\documentclass{emulateapj}

\def\deg{\ifmmode^\circ\else$^\circ$\fi}

\def\arcs{\ifmmode {''}\else $''$\fi}
\def\arcm{\ifmmode {'}\else $'$\fi}
\def\parcs{\sa=.07em \sb=.03em
     \ifmmode $\rlap{.}$^{\scriptscriptstyle\prime\kern -\sb\prime}$\kern -\sa$
     \else \rlap{.}$^{\scriptscriptstyle\prime\kern -\sb\prime}$\kern -\sa\fi}
\def\parcm{\sa=.08em \sb=.03em
     \ifmmode $\rlap{.}\kern\sa$^{\scriptscriptstyle\prime}$\kern-\sb$
     \else \rlap{.}\kern\sa$^{\scriptscriptstyle\prime}$\kern-\sb\fi}

\def\lya{{\rm Ly}$\alpha$}

\def\han {\mbox{{\rm H}$\alpha$}}
\def\ha{\han}

\def\h0{\mbox{H$_0$}}
\newcommand{\ecs}{\mbox{$\rm erg~cm^{-2} sec^{-1}$}}
\def\spose#1{\hbox to 0pt{#1\hss}}
\def\simlt{\mathrel{\spose{\lower 3pt\hbox{$\mathchar"218$}}
     \raise 2.0pt\hbox{$\mathchar"13C$}}}
\def\simgt{\mathrel{\spose{\lower 3pt\hbox{$\mathchar"218$}}
     \raise 2.0pt\hbox{$\mathchar"13E$}}}
\def\lsim{\rlap{$<$}{\lower 1.0ex\hbox{$\sim$}}}
\def\gsim{\rlap{$>$}{\lower 1.0ex\hbox{$\sim$}}}
\def\kms{km~s$^{-1}$}
\def\cm2{cm$^{-2}$}
\received{2003 August 25}
\begin{document}


\title{The Distribution of \lya-Emitting Galaxies at z=2.3\footnotemark[1]}

\footnotetext[1]{Based on observations obtained at Cerro Tololo
Inter-American Observatory, a division of the National Optical
Astronomy Observatories, which is operated by the Association of
Universities for Research in Astronomy, Inc.\ under cooperative
agreement with the National Science Foundation.}

\author{
Povilas Palunas,\altaffilmark{2,4,5}
Harry I. Teplitz,\altaffilmark{3,4,5} 
Paul J. Francis,\altaffilmark{6,7}
Gerard M. Williger,\altaffilmark{8,4} \and
Bruce E. Woodgate\altaffilmark{4}
}
\affil{}

\altaffiltext{2}{McDonald Observatory, University of Texas, Austin, TX 78712}
\altaffiltext{3}{SIRTF Science Center, Caltech, Pasadena, CA 91125}
\altaffiltext{4}{NASA Goddard Space Flight Center, Greenbelt, MD 20771}
\altaffiltext{5}{CUA Research Associate}
\altaffiltext{6}{Research School of Astronomy and Astrophysics, The 
Australian National University, Canberra, ACT 0200, Australia}
\altaffiltext{7}{Joint appointment with the Department of Physics, Faculty
of Science.}
\altaffiltext{8}{Johns Hopkins University, Baltimore, MD 21218}

\begin{abstract}

We present the detection of 34 \lya\ emission-line galaxy candidates
in a 80$\times$ 80 $\times$ 60 co-moving Mpc region surrounding the
known $z=2.38$ galaxy cluster J2143-4423. The space density of \lya\
emitters is comparable to that found by Steidel et~al. when targeting
a cluster at redshift 3.09, but is a factor of $5.8 \pm 2.5$ greater
than that found by field samples at similar redshifts.

The distribution of these galaxy candidates contains several 5-10 Mpc
scale voids. We compare our observations with mock catalogs derived
from the VIRGO consortium $\lambda$CDM n-body simulations. Fewer than
1\% of the mock catalogues contain voids as large as we observe. Our
observations thus tentatively suggest that the galaxy distribution at
redshift 2.38 contains larger voids than predicted by current models.

Three of the candidate galaxies and one previously discovered galaxy
have the large luminosities and extended morphologies of ``\lya\
blobs''.

\end{abstract}


\section{Introduction}

The two-point correlation coefficient of high redshift ($z>2$)
galaxies is quite similar to that of galaxies today (e.g. Steidel
et~al.\ 1996, 2000, Giavalisco et~al. 1998). The {\em topology} of the
distribution of high redshift galaxies, however, is not yet clear.
The two-point correlation coefficient alone is not very sensitive to
this topology. In the local universe, a large fraction of galaxies lie
in filaments and sheets, such as the Great Wall (Geller \& Huchra
1989). These filaments may be several hundred Mpc in length, and are
separated by voids which can be tens of Mpc across.  Our understanding
of how and when filaments and voids were established depends on the
geometry of the universe and biasing of galaxies (e.g. White \& Frenk
1991; Kauffmann, Nusser, \& Steinmetz 1997; Mo, Mao, \& White 1999).

Mapping such large scale structure at high redshifts requires a
careful balance of field of view (to encompass the structures),
redshift coverage (to accept enough test objects in the structure
while avoiding confusion by overlapping structures in distance), and
photometric sensitivity (to find enough test objects). Photometric redshifts
(see Hogg et~al. 1998 for a review) provide an efficient estimate of
redshift for large imaging surveys, but they are not accurate enough
to avoid confusion from multiple overlapping structures, and cannot
delineate the structures directly.  Bright QSOs may be markers of
dense regions (Ellingson, Yee \& Green, 1991), but are too sparse to
provide multiple samples within individual structure elements.  

Narrow-band imaging to find \lya\ emitting galaxies has many
advantages as a technique for mapping the topology of the galaxy
distribution at high redshifts. The technique picks out galaxies in a
narrow-range of redshifts, so the two-dimensional distribution of
their positions on the sky can be used to constrain the topology
alone, without the need for expensive follow-up spectroscopy. At
sufficiently low flux limits the space density of detectable \lya\
emitters becomes comparable to that of Lyman-break galaxies (eg. Hu \&
McMahon 1996).  They should be dense tracers of large scale
structure. With large-format prime-focus imagers, it is relatively
straightforward to survey very large volumes of the early universe.
The fast beam required for large field coverage puts a lower limit the
width of the narrow-band filter. Typically this limits the
line-of-sight depth covered in an exposure to be roughly equal to or
greater than the width.  A disadvantage of \lya\ searches is that the
\lya\ emission from a galaxy is very hard to predict theoretically,
due to the very high optical depth in this line, its dependency on
metalicity and star formation rate, and the ease with which it can be
obscured by dust.

Two studies find some evidence for filamentary structure in the
distribution of \lya\ emitting galaxies on scales of around 5Mpc
(Campos et~al. 1999, M{\o}ller \& Fynbo 2001). Other surveys have
found evidence for clustering in \lya\ emitting galaxies, but not for
filamentary structure (Steidel et~al. 2000, Ouchi et~al. 2002).

In this paper, we present the detection of 34 candidate \lya\ emitting
galaxies in a region containing J2143-4423, a z=2.38 cluster of
galaxies and damped \lya\ absorbers (Francis et~al., 1996, 1997,
2000). Our survey covers a region 80$\times$80$\times$60 co-moving Mpc in size,
and hence is sensitive to much larger structures than
previous studies at comparable redshifts.  The narrow-band imaging
also detects four extended nebulae of \lya\ emission, often dubbed
``blobs'' (Steidel et~al., 1999, Keel et~al. 1999, Francis et~al.,
2000).  Throughout the paper, we assume a $\Lambda$-dominated flat
Universe ($H_0 = 65$ km~s$^{-1}$~Mpc$^{-1}$, $\Omega_M=0.3, \Omega_{\Lambda}=0.7$).

\section{Observations and Data Reduction}

The data were taken with the MOSAIC II instrument on the Blanco 4m
telescope at CTIO.  Observations were made on the nights of 7-8
August, 1999, in the Johnson $UBV$\ and Cousins $RI$\ filters, and in
a custom narrow-band interference filter (NB) centered at 4107\AA\
with a full width at half maximum (FWHM) of 54\AA.  The NB filter was
designed to image \lya\ emission at z=2.38 (Francis et~al. 2000). The
delivered filter covers \lya\ redshifts between $2.356 < z < 2.401$.
The Mosaic II is a prime-focus camera consisting of 8 CCDs, each with
2048$\times$4096 pixels.  The camera has a plate scale of
0.27\arcsec/pixel and covers an overall area of (36\arcmin)$^2$ with
small 9-13\arcsec\ gaps between the CCDs.  Individual exposures were
dithered by 10\arcm . The full NB integration time of 19,800 sec was
achieved for a 26\arcmin$\times$26\arcmin\ region centered on the
J2143-4423 cluster, while shorter integrations extend over a
48\arcmin$\times$50\arcmin\ field. The final trimmed images are
45.9\arcmin$\times$45.5\arcmin\ in size excluding
9\arcmin$\times$9\arcmin\ in the South-East corner with only one
exposure.  The observations are summarized in Table \ref{tab_obs}.


\begin{deluxetable}{llllr}
\tablecaption{Observations\label{tab_obs}}
\tablewidth{0pc}
\tablehead{
\colhead{Filter} & 
\colhead{exptime/frame} &
\colhead{no. frames} &
\colhead{FWHM} &
\colhead{$5\sigma$\tablenotemark{1}} \\
\colhead{} & 
\colhead{(s)} & 
\colhead{} &
\colhead{(\arcs)} &
\colhead{(AB mag.)} 
}
\startdata

$U$  &   1800       &    5   &  1.49   &  23.9   \\
$B$  &    600       &   12   &  1.37   &  26.2   \\
$V$  &    600       &   12   &  1.37   &  25.3   \\
$R$  &    600       &    5   &  1.64   &  24.0   \\
$I$  &    600       &    5   &  1.11   &  23.8   \\
$NB$ &   1800,2700  &    5,4 &  1.35   &  23.5   \\

 \enddata
\tablenotetext{1}{The $5\sigma$\ limit for a point source, on the AB magnitude system.}

\end{deluxetable}

This was the first science run with the MOSAIC II.  The readout
clocking was still being tuned and biases had a slowly varying
harmonic pattern due to the beating of unsynchronized clocks. An FFT
based algorithm was developed to subtract the ringing component of the
bias.  The period of the pattern was approximately 160 pixels
(120\arcsec) in the readout direction on the CCD.  Because this scale
is significantly larger than the size of any source in the images the
subtraction algorithm should not affect the detection or photometry of
of the sources.  All detections were verified in the raw images.

Further image reductions were performed using the {\sc mscred}
package in IRAF\footnote{IRAF is distributed by NOAO, which is
operated by AURA Inc., under contract to the NSF.}.  These include a
correction for crosstalk between the CCDs, bias subtraction and
flat-fielding. 

Twilight flats were used for all of the images except for the
I-band. The I-band twilight flats exhibited considerable fringing with
a different pattern than that due to the night sky. The I-band dome
flats however were not uniformly illuminated and an illumination
correction was derived from highly smoothed twilight flats. The I-band
images were defringed by combining all of the images with object
rejection. The objects were rejected by making a mask using the {\sc
SExtractor} (Bertin \& Arnouts 1996) object images. The object mask
images were convolved with a 20 pixel diameter top hat kernel to
reject the low surface brightness tails of the objects.
 
Wide-field optical distortions were corrected using a standard
astrometric solution for MOSAIC II provided in IRAF.  Final alignment
of the images was performed by matching each image to approximately
2000 stars in the APM catalog.  The alignment was good to 0.3 pixels
(0.08\arcsec) rms.

Before combining the images we created bad pixel masks which included
hot pixels bad columns and cosmic rays. Cosmic rays were selected
using the $jcrrej2$ package in IRAF (Rhoads 2000). We found this
package to work extremely well for finding individual cosmic rays,
however we had difficulty balancing the parameters to eliminate the
faint halos associated with the cosmic ray hits. To eliminate these we
devised a filtering algorithm which calculated the Tookey biweight
statistic (Beers, Flynn \& Gebhardt 1990) in a region around each CR
and eliminated all points adjacent to the CR which were $>2.5\sigma$
from the estimated background level.

\begin{figure}
\plotone{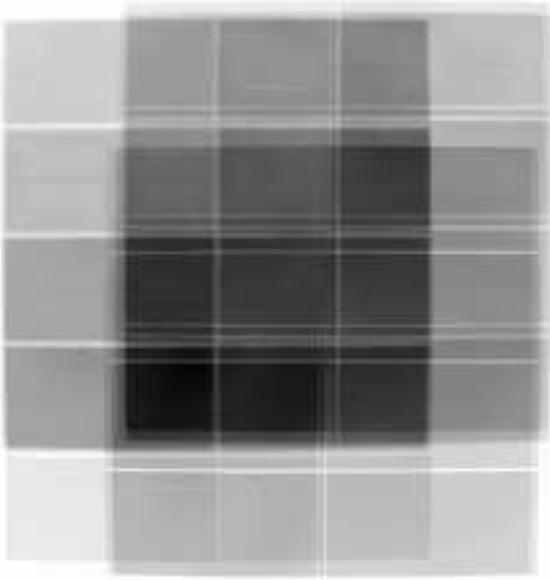}
\caption{Exposure map for the $NB$ imaging. North is up East is to the left. 
\label{fig_expmap}
}
\end{figure}

Photometric calibration was performed using 56 stars in the Landolt
fields 90, 92, SA110 and 113 (Landolt 1992).  A small correction was
include in the calibration to account for Galactic extinction
(Schlegel, Finkbeiner \& Davis 1998).  Catalogs of individual sources
were compiled using {\sc SExtractor}.  The limiting magnitudes vary
across the images. Figure \ref{fig_expmap} shows an exposure map for
the $NB$ imaging. The map was created by summing a set of the flat
field images scaled by exposure times and with the same offsets as the
$NB$ exposures.  The 5$\sigma$ detection limits for point-like objects
in each color are summarized in Table \ref{tab_obs}.  The $NB$ filter
was used as a primary detection band, with measurements made in other
bands using the same apertures, in order to find objects with weak
continua but strong lines. Areas that are covered by only one
exposure, such as the south-east corner are excluded from the
analysis. Photometry was measured in the Kron aperture (Kron 1980),
defined as 2 times the first moment of the radial light
distribution. The first moment is approximately equal to the half
light radius of the distribution. The photometric error was measured
by SExtractor.  Table 1 lists the $5\sigma$\ photometric sensitivity
for a point source in each band.

\section{Results}

We detect 2450 point-like objects and 7000 resolved objects above the
$5\sigma$\ limit in the NB filter.  \lya\ emitters can be identified
in this sample by excess flux in the NB filter compared to the
continuum $B$\ filter.  The limits for $NB-B$\ color excess were
determined by modeling the distribution of photometric errors in the
$NB-B$\ color as a function of the $NB$\ magnitude (see Teplitz
et~al.\ 1998).  The outer envelope of $5\sigma$\ errors that defines
the detection limit for emission-line candidates is fit by the
function:

\begin{equation}
[(NB-15.25)/7.6]^4+0.15
\end{equation}

However, bright emission-line candidates with low equivalent width
(EW) are more likely to be low redshift interlopers than cluster
members (see section \ref{sec_contam}).  We consider candidates above
the $EW=125$\ \AA\ limit to be likely \lya\ emitters.  This limit is
obtained as a broad- minus narrow-band color through the
relation:
\begin{equation}
NB-B = -2.5 \ log {1+EW/W_{B} \over 1+EW/W_{NB}}
\end{equation}
where $W_{B}$ and $W_{NB}$ are the widths of the broad- and
narrow-band filters.

\begin{figure}
\plotone{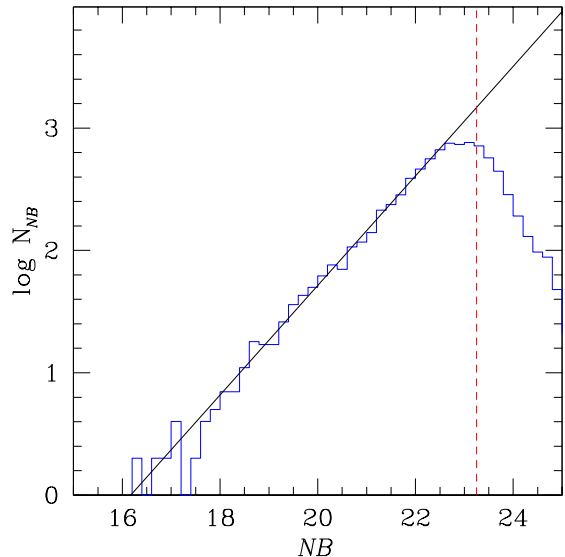}
\caption{Number Counts for resolved objects detected in the
narrow-band images.  Assuming the number counts can be extrapolated by
a power law the detections are 50\% complete at NB=23.25
\label{fig: completeness}
}
\end{figure}

In addition, objects that are only marginally detected in the
narrow-band are unlikely to be good candidates.  For simplicity, we
establish a uniform narrow-band magnitude cut rather than base the cut
on the varying depth across the image.  The magnitude cut is based on
the 50\% completeness limit for narrow-band detections (see Figure
\ref{fig: completeness}).

\begin{figure}
\plotone{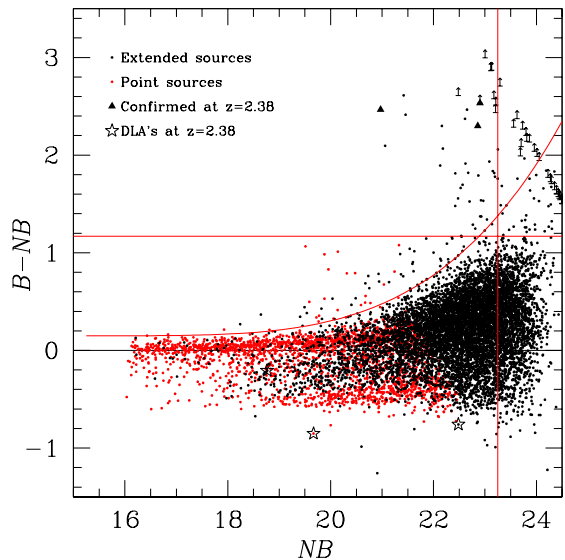}
\caption{The excess flux in the NB filter as a function of magnitude.
Black points are extended sources and grey are unresolved sources.
The large filled circles are the three spectroscopically confirmed
\lya-emitters (Francis et~al.\ 1997).  Upward pointing arrows are
objects undetected in the B-band filter.  Three damped \lya\ absorbers
(Francis \& Hewett 1993, Francis, Wilson \& Woodgate 2001) are
indicated by open stars.  The curved lines indicated the $5\sigma$\
confidence level for excess emission; that is, points above the line
are candidate emission-line object.  The horizontal line is the
$EW_{obs}=125$\ \AA\ level.  The vertical line is the 50\%
completeness cutoff; that is, only candidates brighter than that line
are considered likely to be real. Objects which are simultaneously
above the $5\sigma$\ threshold, above the observed $EW_{obs}=125$\
\AA\ level and brighter than the 50\% completeness limit are selected
as \lya-emission candidates.
\label{fig: nb-b excess}}
\end{figure}

Figure \ref{fig: nb-b excess}\ shows the $NB-B$\ color excess for all
objects detected in the $NB$\ image.  The \lya\ emitters from Francis
et~al.\ (1996) are clearly detected in the new $NB$\ measurement.  For
new emission line candidates we consider objects that are detected
above the 50\% completeness limit in the narrow-band filter ($NB<
23.5$), are located above the $5\sigma$\ color excess line, and have
an observed EW greater than 125 \AA.  We can detect galaxies with
emission line strengths brighter than $1.4 \times 10^{-16}$\ ergs
cm$^{-2}$\ s$^{-1}$.  This flux limit corresponds to a \lya\
luminosity of $1.9 \times 10^9 L_{\sun}$. Thirty-seven spatially
resolved $NB$\ excess objects are detected.  Table \ref{tab_nbdet}
lists the positions and measured properties of these objects.

\begin{deluxetable*}{rrlllll}
\tablecaption{Emission-Line Galaxies \label{tab_nbdet}}
\tablewidth{0pc}
\tablehead{
\colhead{$\alpha$\ (2000)} & 
\colhead{$\delta$\ (2000)} &
\colhead{$NB$} &
\colhead{$B-NB$} &
\colhead{\lya\ flux} &
\colhead{$EW_{0}$} &
\colhead{Notes}\\
\colhead{} & 
\colhead{} &
\colhead{(mag.)} & 
\colhead{(mag.)} &
\colhead{(\ecs)} &
\colhead{(\AA)} &
\colhead{} 
}
\startdata

  21:40:19.98 & -44:19:48.2 & 22.62 & 1.91  &  2.73$\times10^{-16}$  &  380   &    \\ 
  21:40:31.00 & -44:36:04.5 & 22.97 & 1.53  &  1.81$\times10^{-16}$  &  215   &    \\ 
  21:40:33.10 & -44:36:10.8 & 22.91 & 1.37  &  1.81$\times10^{-16}$  &  169   &    \\  
  21:40:36.77 & -44:20:28.8 & 23.20 & 1.78  &  1.56$\times10^{-16}$  &  312   &    \\ 
  21:40:48.09 & -44:31:01.7 & 23.11 & 2.87  &  1.95$\times10^{-16}$  & 3026   &    \\ 
  21:40:48.97 & -44:01:23.6 & 22.77 & 1.29  &  2.00$\times10^{-16}$  &  150   &    \\ 
  21:40:58.22 & -44:00:22.0 & 23.00 & 3.00  &  2.18$\times10^{-16}$  & 5900   &    \\ 
  21:41:02.90 & -44:01:55.9 & 22.81 & 1.25  &  1.89$\times10^{-16}$  &  141   & \\ 
  21:41:07.38 & -44:38:11.7 & 23.04 & 1.34  &  1.59$\times10^{-16}$  &  162   &    \\ 
  21:41:44.41 & -44:37:06.7 & 22.68 & 1.82  &  2.54$\times10^{-16}$  &  331   &    \\ 
  21:41:47.66 & -44:21:21.9 & 22.61 & 2.37  &  2.95$\times10^{-16}$  &  824   &    \\ 
  21:41:53.55 & -44:38:18.3 & 23.18 & 1.82  &  1.60$\times10^{-16}$  &  331   &    \\ 
  21:42:06.03 & -44:34:47.9 & 23.18 & 2.58  &  1.79$\times10^{-16}$  & 1277   & B6 \\ 
  21:42:14.28 & -44:32:15.8 & 22.18 & 2.07  &  4.21$\times10^{-16}$  &  489   &    \\ 
  21:42:27.56 & -44:20:30.1 & 20.97 & 2.47  &  1.35$\times10^{-15}$  & 1004   & B1 \\ 
  21:42:28.54 & -44:32:38.5 & 23.20 & 2.44  &  1.73$\times10^{-16}$  &  944   &    \\ 
  21:42:29.73 & -44:21:02.8 & 22.86 & 2.30  &  2.33$\times10^{-16}$  &  723   & B2 \\ 
  21:42:32.20 & -44:20:18.6 & 22.90 & 2.53  &  2.30$\times10^{-16}$  & 1141   & B4 \\ 
  21:42:34.88 & -44:27:06.2 & 21.80 & 3.00  &  6.95$\times10^{-16}$  & 5900    & B7 \\
  21:42:42.63 & -44:30:09.0 & 21.20 & 3.00  &  1.14$\times10^{-15}$  & 5900   &    \\ 
  21:42:54.07 & -44:14:39.7 & 22.93 & 1.37  &  1.78$\times10^{-16}$  &  169   &    \\ 
  21:42:56.34 & -44:37:56.8 & 22.48 & 1.56  &  2.86$\times10^{-16}$  &  225   &    \\ 
  21:43:00.09 & -44:19:21.7 & 22.57 & 1.70  &  2.73$\times10^{-16}$  &  277   &    \\ 
  21:43:03.57 & -44:23:44.2 & 21.42 & 2.61  &  9.06$\times10^{-16}$  & 1371   & B5 \\ 
  21:43:03.80 & -44:31:44.9 & 22.16 & 2.30  &  4.43$\times10^{-16}$  &  723   &    \\ 
  21:43:05.90 & -44:27:21.0 & 21.06 & 2.10  &  1.19$\times10^{-15}$  &  513   &    \\ 
  21:43:06.42 & -44:27:00.6 & 22.48 & 2.61  &  3.41$\times10^{-16}$  & 1371   &    \\ 
  21:43:11.48 & -43:59:01.0 & 23.13 & 2.87  &  1.91$\times10^{-16}$  & 3026   &    \\ 
  21:43:22.22 & -44:13:06.5 & 23.09 & 1.93  &  1.78$\times10^{-16}$  &  392   &    \\ 
  21:43:23.80 & -44:41:36.4 & 22.87 & 1.79  &  2.12$\times10^{-16}$  &  317   &    \\ 
  21:43:24.06 & -44:27:59.9 & 22.42 & 1.81  &  3.22$\times10^{-16}$  &  326   &    \\ 
  21:43:37.41 & -44:17:53.4 & 23.21 & 2.51  &  1.73$\times10^{-16}$  & 1092   &    \\ 
  21:43:37.48 & -44:23:52.8 & 22.65 & 2.47  &  2.88$\times10^{-16}$  & 1004   &    \\ 
  21:43:44.92 & -44:05:46.3 & 22.53 & 1.66  &  2.81$\times10^{-16}$  &  261   &    \\ 
  21:43:48.30 & -44:08:26.9 & 22.42 & 1.48  &  2.95$\times10^{-16}$  &  200   &    \\ 
  21:44:12.15 & -44:05:46.6 & 21.46 & 2.41  &  8.56$\times10^{-16}$  &  890   &    \\ 
  21:44:12.97 & -43:57:56.7 & 22.70 & 1.53  &  2.31$\times10^{-16}$  &  215   &    \\

\enddata

\end{deluxetable*}

In addition, 7 unresolved objects with $UBV$ colors consistent with QSO's
(e.g., Hall et~al. 1996a, 1996b) are detected in emission with
equivalent widths greater than 30\AA\ in the observed frame. Table
\ref{tab_nbqsos} lists the positions and measured properties of these
objects. 

Finally, a QSO at z=2.38 was found by Hawkins (2000).  We detect this
object in emission, but it was not selected as a QSO candidate because
the $UBV$ colors are within the stellar locus. We include this QSO in
Table \ref{tab_nbqsos}.

\begin{deluxetable*}{rrlllllll}
\tablecaption{Emission-Line QSO Candidates \label{tab_nbqsos}}
\tablewidth{0pc}
\tablehead{
\colhead{$\alpha$\ (2000)} & 
\colhead{$\delta$\ (2000)} &
\colhead{$NB$} &
\colhead{$B-NB$} &
\colhead{line flux} &
\colhead{$EW_{0}$} &
\colhead{$B-V$} &
\colhead{$U-B$} &
\colhead{Notes}\\
\colhead{} & 
\colhead{} &
\colhead{(mag.)} & 
\colhead{(mag.)} &
\colhead{(\ecs)} &
\colhead{(\AA)} &
\colhead{(mag.)} & 
\colhead{(mag.)} &
\colhead{} 
}
\startdata
    21:40:52.46 &  -44:36:21.35 & 19.96 & 0.83 & 2.04e-15 &  70 &  0.26 & -0.28 & \\
    21:41:54.50 &  -44:18:35.97 & 20.14 & 1.01 & 1.96e-15 &  96 &  0.05 & -0.31 & z=1.66 \\
    21:42:35.12 &  -44:32:30.36 & 21.35 & 0.99 & 6.35e-16 &  93 &  0.36 & -0.11 & \\
    21:42:43.49 &  -44:14:25.32 & 20.14 & 0.41 & 1.02e-15 &  27 &  0.07 & -0.35 & \\
    21:43:16.44 &  -44:16:51.54 & 21.04 & 0.67 & 6.51e-16 &  51 &  0.08 & -0.30 & \\
    21:43:20.38 &  -44:20:00.35 & 20.59 & 0.46 & 7.39e-16 &  31 &  0.06 & -0.52 & \\
    21:43:22.53 &  -44:31:49.18 & 19.88 & 0.98 & 2.45e-15 &  91 &  0.04 & -0.43 & \\
    21:43:26.25 &  -44:26:03.44 & 20.77 & 0.49 & 6.58e-16 &  34 &  0.26 & -0.33 & \\

\enddata

\end{deluxetable*}

\subsection{Foreground Contamination}\label{sec_contam}

For spatially resolved narrow-band excess sources the only likely
contaminating line is [O~II] (3727\AA ).  Our narrow-band filter
centered at 4107\AA, is sensitive to [O~II] emission from foreground
$0.095 < z < 0.109$ galaxies. Indeed, we detect emission from many
well resolved spiral galaxies below our EW cutoff.

[O~II] interlopers within our \lya\ sample would have absolute B magnitudes
around $-13.8$, and would hence be Blue Compact/HII galaxies (BCGs).
At this redshift they would only be marginally spatially resolved, so we could
not separate them from $z=2.38$ \lya\ emitters by morphology alone.

To be detected through our filter, they must lie within a box of comoving 
size 5.9$\times$5.9$\times$60 Mpc, and have a rest-frame [O~II] equivalent
width greater than 114\AA . Using the local luminosity function of Jerjen, 
Bingelli \& Freeman (2000) we would expect to find $\sim 12$ BCG 
galaxies in such a box. Pustilnik et~al. (1999), however, found that only 
$\sim 8$\% of BCG galaxies have [O~II] equivalent widths large enough to meet
our selection criteria. If this foreground region was an average one, we would
therefore expect $\sim 1$ foreground BCG galaxy to be
contaminating our sample.

\begin{figure}
\plotone{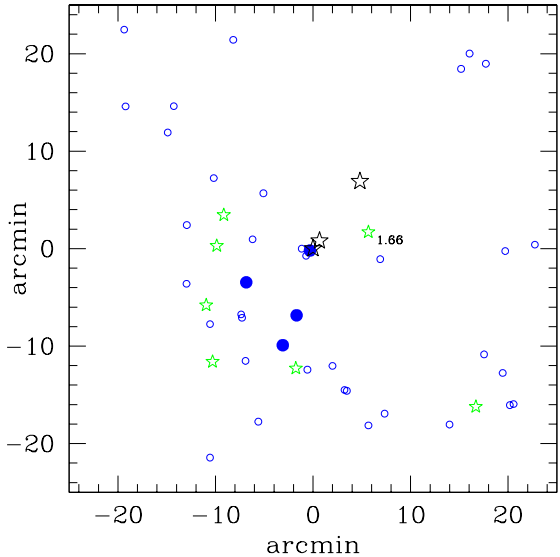}
\caption{The distribution of emission-line objects in the field.  The
small open circles represent emission-line galaxy candidates.  The
large filled circles are extended emission-line nebulae (blobs).  The
large open stars are the three known DLAs, and the small open stars are
quasar candidates at the cluster redshift or lower redshifts.  East is
to the left and North is upwards. \label{fig: lya positions} }
\end{figure}

The southernmost 80\% of our field was included in the Las Campanas Redshift
Survey (LCRS, Shectman et~al. 1996), which has excellent sensitivity to large 
scale structures at redshift $z \sim 0.1$. The region in which foreground
[O~II] emitters could contaminate our survey contains a single isolated
cluster (Abell 3800, Abell, Corwin \& Olowin 1989), but is otherwise empty of
galaxies. The LCRS detects many filaments of galaxies at this redshift, but
none lie in our field at redshifts that could cause [O~II] emission to
impersonate \lya\ at z=2.38. Despite the presence of Abell 3800, the 
galaxy density is indeed slightly below the LCRS average at this
redshift. Abell 3800 is located 10\arcmin\ west and 3\arcmin\ south of our
field center (see Fig~\ref{fig: lya positions}). We see no concentration of
candidate \lya\ emitting galaxies at this location: indeed quite the
opposite. Only one of our candidates lies within one projected Mpc of 
this location. Three others lie around 1.3 projected Mpc from the cluster, but
these are the three for which we have spectroscopic confirmation that they
lie at $z=2.38$.  We therefore tentatively conclude that
foreground contamination is unlikely to be a big problem.

Foreground contamination is more likely a problem for the 7
narrow-band excess QSO candidates. The narrow-band excess could be
caused by \lya\ emission at redshift 2.38, but could also be quasars
at other redshifts, with different lines producing the excess. Indeed
one of these sources was found (Francis et~al. 1997) to be a QSO at
z=1.66, with its C~IV (1549\AA ) emission producing the narrow-band
excess.  The luminosity and redshift distribution of QSOs implies that
about 60\% or 4 of these quasars should be \lya\ emitters at redshift
2.38 (see Palunas et~al. 2001). If some of these sources are QSOs at
redshift 2.38, it will be further evidence that QSOs lie in the most
massive dark matter halos, which in turn exist in the most highly
clustered environments (e.g. Silk \& Weinberg 1991).

\section{Discussion}

\subsection{The Space Density of \lya\ Emitters}

\begin{figure}
\plotone{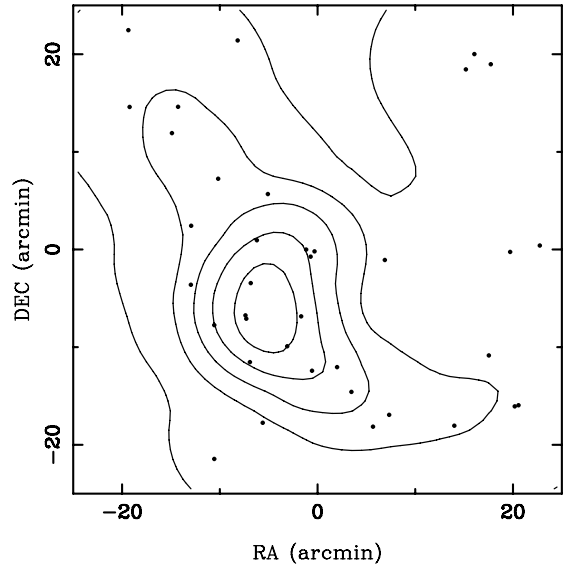}
\caption{Density contours for the distribution of \lya\ candidates
estimated with the adaptive kernel routine. The lowest contour is one
galaxy per resolution element $3.3\times10^{-3}$ galaxies
arcmin$^{-2}$. The highest contour is $5.7\times10^{-2}$ galaxies
arcmin$^{-2}$.\label{fig_lyacont}}
\end{figure}

We are sensitive to \lya\ emitting galaxies over a
45\arcmin$\times$45\arcmin\ field minus 9\arcmin$\times$9\arcmin in
the South-East corner, in the redshift range $2.356 < z < 2.401$.  For
our adopted cosmology, this corresponds to a volume 6100 co-moving
Mpc$^2$ in area and 59 co-moving Mpc deep.
Figure~\ref{fig_lyacont} shows the density contours of \lya\
candidates estimated using an adaptive kernel routine (Beers, Flynn \&
Gebhardt 1990).  The initial smoothing scale is 12 arcminutes. The
highest density contour is $5.7 \times 10^{-2}$ sources per square
arcminute corresponding to a spatial density of $3.0 \times 10^{-4}$
sources per cubic Mpc. The average density over the whole field is $1.9
\times 10^{-2}$ sources per square arcminute corresponding to a
spatial density of $1 \times 10^{-4}$ sources per cubic Mpc.

Steidel et~al. (2000) surveyed 20,000 cubic co-moving Mpc at redshift
3.09, targeting a cluster of Ly-dropout galaxies. Their survey was
sensitive to lower \lya\ luminosities than ours, and they used a lower
rest-frame equivalent width threshold. They find 9 candidates meeting
our luminosity and equivalent width thresholds corrected for redshift,
yielding a density of $4.5 \times 10^{-4}$ sources per cubic
Mpc. Steidel et~al. calculate that this space density is a factor of
$6.0 \pm 2.4$ greater than that of smaller survey of Cowie \& Hu
(1998). This implies that peak density in our field is a factor of 4
overdense.



\subsection{Star Formation Rates}
 
To estimate the star formation rate we first assume that star
formation powers the observed \lya\ flux.  This is almost certainly a
great underestimate: even tiny amounts of dust will greatly reduce the
observed \lya\ flux due to its high optical depth.  We use an
un-reddened \lya/\ha\ ratio of 8:1, and the conversion factor between
H$\alpha$ flux and star formation rate of Kennicutt (1983). Our
faintest candidates (NB=23.25) have inferred star formation rates of
$\sim 9 M_{\sun}{\rm yr}^{-1}$.

Summing over all galaxies down to our flux limit, and making a crude
correction for incompleteness amongst the faintest, the integrated
star formation rate density (assuming no dust obscuration) is $2.4
\times 10^{-3} M_{\sun}{\rm Mpc}^{-3}{\rm yr}^{-1}$: well below the
value ($\sim 0.1 M_{\sun}{\rm Mpc}^{-3}{\rm yr}^{-1}$) derived from
integrated blue light by Madau, Pozzetti \& Dickinson (1998).

We also estimate star formation rate from the 1600\AA\ UV continuum
using the Madau et~al. formula. At z=2.38 the 1600\AA\ continuum falls
in the $V$-band. We detect 23 of our candidates in the V-band.  The
star-formation density implied by these candidates is $5 \times
10^{-3} M_{\sun}{\rm Mpc}^{-3}{\rm yr}^{-1}$.  The star formation rate
for a candidate at the V-band limit of 25.3mag is $\sim 18
M_{\sun}{\rm yr}^{-1}$.  The 14 galaxies not detected in the $V$-band
could add, as an upper limit, $0.6 \times 10^{-3} M_{\sun}{\rm
Mpc}^{-3}{\rm yr}^{-1}$ to the star formation rate density.

Our \lya\ candidates therefore contribute only a small fraction ($\sim
5\%$) of the overall star formation rate density at z=2.38.

In a detailed study of the spectra of Lyman-break galaxies, Shapley
et~al. (2003) conclude that galaxies with the strongest \lya\ emission
have bluer UV continua, weaker interstellar absorption and smaller
star-formation rates than galaxies with weak \lya\ emission or \lya\
absorption. \lya\ emission from galaxies with the highest
star-formation rates is absorbed by dust.

\subsection{The Distribution of \lya\ Emission Candidates}

The distribution of our candidate resolved \lya\ emitting sources is
shown in Fig~\ref{fig: lya positions}. There is no obvious overdensity
surrounding the previously known cluster at the field center. Instead,
most of the candidates lie in a broad band extending from the
north-east of the field to the south.  There is a significant is the
lack of galaxies in the north and west. These regions have long
integration times. There are three close pairs or triplets of
candidates, but otherwise the candidates are well spaced.  The pairs
have a spacing of about 20 arcsec or a minimum separation of 200
proper kpc.  The spacing of the galaxies in the triplet is about 50
arcsec or a minimum separation of 440 proper kpc.

\subsubsection{Comparison to a random distribution}

We used two statistics to compare our data to a random distribution.
We compare these statistics to results from Monte Carlo simulations
generated with the same exposure-time mask. In this section we show
that the distribution of our candidates has a significant excess of
close ($< 1$\arcmin ) pairs, and of large (6\arcmin\ -- 8\arcmin )
voids.

The first statistic is essentially the angular two-point correlation
function. For a series of angular separations, we take each data point
in turn and count the number of other data points within that angular
radius.

The second statistic is a two-dimensional analogue of the void
probability function (VPF). For each angular scale, we randomly place
1000 circles with that radius on our field, so that the circles do not
extend beyond the edge of our data. We count the fraction of these
circles which did not contain any data points. The requirement that
the circles not extend beyond the edge of the data makes this
statistic most sensitive to voids near the center of the field. Void
probability functions are sensitive to structures such as voids and
filaments, to which the two-point correlation function is
relatively insensitive.

Our data were compared to random Monte Carlo simulations generated as
follows.  At each location in our image, the total exposure time was
calculated. Using these total exposure times, the relative flux limits
in every part of our field were computed. The number/magnitude
distribution of our brighter candidates can be reasonably approximated
as a power-law, with the number per unit magnitude increasing by a
factor of roughly 5. Combining this number-magnitude
relation with the flux limits, we calculated the relative probability
of finding a candidate per unit area in any given part of our
image. We then used this relative probability map to generate 2000
fake data sets, each containing the same number of sources as in the
real data.

Our data have a significant excess of galaxies with separations less
than 1\arcmin . Only 0.1\% of the Monte Carlo simulations produced
this many close galaxies. On larger scales, our data still show an
excess of galaxies, but this is no longer significant at the 5\%
level.

\begin{figure}
\plotone{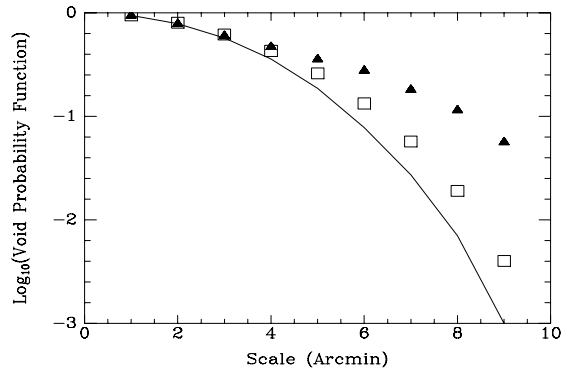}
\caption{
The void probability function of our data (solid triangles) compared to the
average void probability functions of randomly distributed points (solid line)
and void probability functions derived from the $\lambda$CDM n-body
simulations (open squares).
\label{fig: vpf}
}
\end{figure}

The void probability function (Fig~\ref{fig: vpf}) is more
interesting.  There is a clear excess of voids on scales of 5 arcminutes
or greater. On scales of 5\arcmin\ -- 8\arcmin , this excess is
significant at the 99.9\% confidence level (ie. fewer than 0.1\% of
our randomly generated data sets had VPFs as large as our data).

We conclude that there is a significant excess of voids, and of small
scale clustering, in our data. This conclusion is insensitive to the
details of how we generated the random data sets: changes in (for
example) the assumed number/magnitude relation made little difference
to our results.

\subsubsection{Comparison with $\lambda$CDM n-body Simulations}

The distribution of our \lya\ candidates is thus inconsistent with a
random distribution, with 99.5\% confidence. Assuming that they are indeed
galaxies at redshift 2.38, is their distribution consistent with theoretical
predictions? In this section, we show that while the two-point correlation
coefficient of our galaxy candidates is quite consistent with the predictions
of a $\lambda$CDM simulation, the distribution of our candidates shows more 
voids on 6\arcmin\ -- 8\arcmin\ scales, with 97\% confidence.

To evaluate the apparent excess of voids, we compared our measured
distribution of galaxies against the $\lambda$CDM n-body simulations
performed by the VIRGO consortium (Kauffmann et~al. 1999). We used
their galaxy catalog, generated for redshift 2.12. For each galaxy,
they list its position, stellar mass, gaseous mass and star formation
rate.

Mock catalogs were generated as follows. Our observations were
targeted at the source B1, a probable pair of giant elliptical
galaxies surrounded by an extensive neutral hydrogen cloud (Francis
et~al. 2000). B1 has an estimated stellar mass of $10^{10.9} M_{\sun}$
and gaseous mass of $10^{9.7} M_{\sun}$. We picked five galaxies in
the $\lambda$CDM simulations that had stellar and gaseous masses at
least this large, and which lay far enough from the edge of the
simulation volume.  The regions around these ``B1-analogue'' galaxies
should be good matches to our data. For comparison, we also picked
eight random locations within the $\lambda$CDM simulation volume. We
then extracted the galaxies that lay within a volume
70$\times$70$\times$46 co-moving Mpc (for our adopted cosmology)
centered at each of these locations. There were $180 \pm 30$ (1
$\sigma$) galaxies in their catalogs in each of these volumes.

We then had to sparsely sample the galaxies within each of these volumes to
generate mock catalogs with (a) 37 galaxies, and (b) a probability of being
detected consistent with our exposure-time map. We did this in three different
ways.

\begin{enumerate}

\item Randomly. As \lya\ emission can be generated in many ways (by
star formation, shocks, cooling flows, AGN etc), and its escape in
measurable quantities depends on accidents of the geometry and dust
distribution inside galaxies, we first assumed that all the galaxies
in the $\lambda$CDM simulations had the same probability (37/180
$\sim$ 20\%) of being detected by our survey.

\item Star Formation Rate. If the \lya\ flux is generated by star
formation, only galaxies with star formation rates in excess of $8
M_{\sun}{\rm yr}^{-1}$ would have been detected. We therefore took
only galaxies with star formation rates at least this large, and
generated random sub-samples with the correct size and probability of
detection as a function of position in our field. Only 25\% of the
galaxies within the $\lambda$CDM cubes had star formation rates this
large.

\item Mass. We took only galaxies with masses greater than $10^{10.3}
M_{\sun}$, once again picking random sub-samples of the appropriate
size and with probabilities of detection scaling appropriately with
exposure time in each part of our image.

\end{enumerate}

For each data cube, and for each of these three sub-sampling
techniques, we repeated the random sub-sampling 2000 times. We then
calculated the angular two-point correlation function and the void
probability function for each sub-sampling of each data cube, and
compared our results with the observations.

\begin{figure}
\plotone{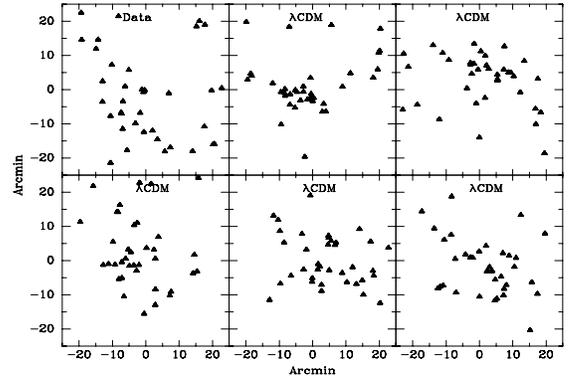}
\caption{
The distribution on the sky or our candidate \lya emitting galaxies (top
left) compared to five simulated data sets drawn from the $\lambda$CDM n-body
simulations. These simulations use random sub-sampling and are centered on
B1-analogue galaxies.
\label{fig: skyplot}
}
\end{figure}

Fig~\ref{fig: skyplot} shows our data, and five of the mock
$\lambda$CDM data cubes (centered on the five B1-analogue
galaxies). The cubes shown have been randomly sub-sampled: those
sampled on the basis of star formation rate or mass look very
similar. Fig~\ref{fig: vpf} shows the average void probability
function of the randomly sub-sampled data cubes (centered on
B1-analogue galaxies).

The angular two-point correlation coefficient of our data, and of the
$\lambda$CDM simulations are in excellent agreement. As Fig~\ref{fig: vpf}
shows, however, the agreement is not as good for the void probability function.
Our data have a void probability function considerably higher than that of the
simulations on large scales.

Is this difference significant? The VPFs of our 13 simulated data
cubes (five centered on B1-analogue galaxies and eight randomly
located) actually showed rather little scatter, when averaged over all
the 2000 random sub-samplings of each. The dominant source of scatter
was the selection of the 37 galaxies in each simulated volume, which
was typically twice as large.  Our uncertainty is thus dominated by
the small sample size rather than by cosmic variance.

If we take the five data cubes centered on B1-analogues, and generate
2000 random sub-samplings of each, fewer that 0.1\% of all these
random sub-samplings have void probability functions as large as we
observe on scales of 5\arcmin\ -- 8\arcmin . If we sub-sample on the
basis of star formation rate or galaxy mass, this fraction rises to
around 1\%. We would expect this, as more massive galaxies, and those
with larger star formation rates, are expected to be more strongly
biased. If we repeat these calculations for the randomly centered data
cubes, the void probability function also rises slightly. This is
presumably because centering our data cube on a B1 analogue precludes
the possibility of a void in the center of the field.

Thus the observed distribution of galaxies has an excess of voids over the
$\lambda$CDM simulations, significant at roughly the 99\% level. A qualitative
impression of this can be gained from Fig~\ref{fig: skyplot}. While all the
$\lambda$CDM simulations show empty regions near the edge of the field, this
is an artifact of the shorter exposure times in these regions. The simulations
do not show large voids in the central $\pm 13$ arcminutes. Due to the restriction
that we only place voids so that they do not overlap the edges of our field,
our VPF statistic is mostly sensitive to the distribution of galaxies in this
central region.

\subsubsection{Discussion of the Voids}

We thus have tentative evidence from our survey of larger voids in the galaxy
distribution at redshift 2.38 than predicted by one particular $\lambda$CDM
simulation. With only 97\% confidence, a sample size of 34 and no spectral 
confirmation for most of our candidate galaxies, this evidence must be
regarded as tentative at best. But if it is confirmed by larger samples, what
is it telling us?

Steidel et~al. (2000) did not see voids, but they covered too small a
region on the sky to be sensitive to the scales of void we are
detecting. Ouchi et~al. (2003) imaged a region $\sim 25$\% the size of
ours, searching for \lya\ candidates at $z=4.86$. Voids can be seen in
their data if only brightest objects are selected so that their number
per unit area matches ours.  Voids might also explain the enormous
field-to-field variance in the space density of \lya\ sources found by
Pascarelle, Windhorst \& Keel (1998).  The spiky distribution of the
redshifts of Lyman-dropout galaxies seen by Steidel et~al. (1998) and
Adelberger et~al. (1998) would also be consistent with large voids in
the high redshift galaxy distribution.

If the distribution of galaxies really contains more 10 co-moving Mpc scale
voids than predicted, either $\lambda$CDM is predicting too uniform a
distribution of galaxies, or the voids contain dark matter halos which for
some reason do not contain detectable galaxies.

There is some evidence for the latter hypothesis. The sight-line
to background QSO 2138$-$4427 passes through one of the voids in our data, but
the QSO spectrum shows a metal enriched damped \lya\ absorption-system 
at our redshift (Francis \& Hewett 1993, Francis, Wilson \& Woodgate 2001). 
So there is clearly at least one dense concentration of gas and some stars 
within one of our voids.

\subsection{Blobs}

Some high redshift galaxy clusters contain extended \lya\ emission
nebulae, or ``blobs'' (Steidel et~al.\ 1999).  The blobs are radio
quiet ($< 140~\mu$Jy at 1.4 GHz), but otherwise share morphological 
similarity with the nebulae
around radio galaxies (e.g. Kurk et~al.\ 2001).  \lya\ blobs are large, 
bright gas
clouds ($\sim 100 h^{-1}$~kpc, $\Omega_m =0.3, \Omega_{\Lambda}=0.7$;
\lya\ flux up to $1.8\times \sim 10^{-15}$ erg cm$^{-2}$ s$^{-1}$).
They appear to be common in regions of significant galaxy overdensity
($\sim 10$~times that of the field) at high redshift.  Most blobs are
associated with a host galaxy, though not symmetrically centered on
it.  Blobs may break into smaller knots of \lya\ and continuum
emission, with velocity differences $\Delta V\simlt 2000$ \kms\
(Steidel et~al. 2000).   Blobs can be among the brightest high-$z$
sub-mm sources (Chapman et~al.  2001).  Some blobs 
(Francis et~al. 2001) show CIV emission (and may host AGN) but others
do not.

\begin{figure}
\plotone{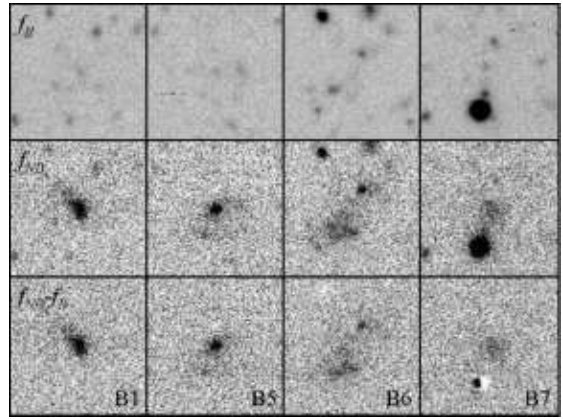}
\caption{REDUCED RESOLUTION. Images of the blobs.  The top row shows
the broad band image of each blob; the middle row the narrow-band
image; and the bottom row the difference of the two, scaled to give an
image of the \lya\ emission.  Each image is $30\arcs \times 30\arcs$.
\label{fig: blob images} }
\end{figure}

We detect four \lya\ blobs in the narrow-band image.  One of these,
``B1'', has previously been detected (Francis et~al.\ 2001).  Figure
\ref{fig: blob images}\ shows the broad- and narrow-band images of
each blob, together with the continuum subtracted emission-line image.
Each blob conforms to the expected size and brightness of this class
of object.  Table \ref{tab: blobs}\ summarizes the characteristics of
the blobs.

\begin{deluxetable*}{llllll}
\tablecaption{\lya\ Blobs \label{tab: blobs}}
\tablewidth{0pc}
\tablehead{
\colhead{Name} & 
\colhead{Redshift} &
\colhead{major axis} &
\colhead{\lya\ flux} &
\colhead{CIV emission} &
\colhead{Ref.\tablenotemark{a}} \\
\colhead{} & 
\colhead{($z$)} & 
\colhead{(\arcs)} &
\colhead{(\ecs)} &
\colhead{} &
\colhead{} 
}
\startdata

SB-1 & 3.09 & 17           & $1.4\times10^{-15}$    & no         & 1,2          \\
SB-2 & 3.09 & 15           & $1.2\times10^{-15}$    & no         & 1,2          \\
KB-A & 2.39 & $\sim$8      & $1.1\times10^{-15}$    & yes        & 3,4          \\
KB-B & 2.39 & $\sim$8      & $0.45\times10^{-15}$   & yes        & 3,4          \\
B1   & 2.38 & 17           & $1.8\times10^{-15}$    & yes        & 5, this work \\
B5   & 2.38 & 8            & $1.5\times10^{-15}$    & \nodata    & this work    \\
B6   & 2.38 & 8            & $1.4\times10^{-15}$    & \nodata    & this work    \\
B7   & 2.38 & 7            & $0.7\times10^{-15}$    & \nodata    & this work    

\enddata
\tablenotetext{a}{
1 -- Steidel et~al. 2000; 2 -- Bautz \& Garmire 2001; 
3 -- Windhorst et~al. 1998; 4 -- Keel et~al. 1999; 
5 -- Francis et~al. 2001}

\end{deluxetable*}

These blobs may be pre-galactic clouds of gas, in the process of
initial collapse or destined to be stripped away into the intracluster
medium (ICM).  In the blobs, we may witness a stage of structure
formation which is illuminated by an unusual or transitory phenomenon.
The excitation mechanism for the \lya\ emission is not known, but
three primary models have been suggested:

In the early stages of a galaxy's formation, gas collapses into the
dark matter potential well and cools via radiation (e.g. Fabian
et~al. 1986).  Haiman, Spaans \& Quataert (2000) calculate that
cooling flows (with gas temperatures of a few $\times 10^4$ K) could
produce the observed flux range, size, surface brightness profiles and
number density observed from \lya\ blobs.  While Fardal et~al. (2000)
doubt that cooling flows would have sufficient luminosity, Francis
et~al.  (2001) suggest that shocks within the infalling gas can
trigger the blob emission in the cooling flow.  Confirmation that the
blobs are high redshift cooling flows would have profound implications
for the interpretation of semianalytic galaxy formation models (e.g.
Fall \& Efstathiou 1980) since we would be observing the initial
collapse of gas into dark matter potential wells.

Taniguchi et~al. (2001 and references therein) suggest that
starburst-powered winds trigger \lya\ emission in the expelled gas.
Dust could obscure the host galaxy, with the \lya\ emission arising
from a superwind extending $\sim 100$ kpc from the host.  The dusty
host galaxies would eventually evolve into ordinary elliptical
galaxies, and the blobs would presumably dissipate into the
intracluster medium.  The SCUBA detections support this model,
implying rapid star formation (Chapman et~al. 2001) and a probability
of hosting an AGN (Barger et~al.  2001).  Superwinds would indicate
that the blobs are the next phase of galaxy evolution after the
initial collapse.  Taniguchi \& Shioya (2000) estimate this phase
would last only 0.1 Gyr, occurring after the first 0.5 Gyr of star
formation.  The expelled gas would eventually be stripped into the
ICM, though it would be metal-enriched having been processed by
supernovae in the host galaxy.

An AGN may photo-ionize the gas, similar to radio galaxy nebulae and
Seyfert 2 galaxies with extended emission line regions
(e.g. Villar-Martin, Tadhunter \& Clark 1997).  For example, a \lya\
nebula has been found around the HDF-S QSO which is radio-quiet
(Bergeron et~al. 1999; Palunas et~al. 2000a).  For the blobs, we do
not see a strong UV continuum source; however, there may be an AGN
surrounded by an obscuring torus (e.g. Krolik \& Begelman 1986). In
fact, most of the Chandra point sources do not show optical evidence
for AGN (Mushotzky et~al. 2000). Assuming an ionization parameter of
10 and a density of 1 cm$^{-3}$\ at a distance of 10 kpc, a central
source with luminosity $9\times 10^{45}$ ergs s$^{-1}$\ is required to
ionize a \lya\ blob with $L_x\sim 10^{44}$~ergs s$^{-1}$.  The gas in
AGN-supported blobs would contribute to the ICM, but with potentially
lower metallicity than that from a starburst, thus diluting rather
than enriching the ICM.

\section{Summary}

Narrow-band imaging in the J2143-4423 region has revealed 34 candidate
\lya\ emitters, including three new extended \lya\ blob candidates and
five possible quasars.  Together with the previously reported
emission-line galaxies and DLA systems, these detections suggest this
field is one of the most highly evolved structures detected at high
redshift.

The distribution of the candidates is non-random. There are more closely
grouped galaxies, and more voids in the distribution, than would be expected
from either a random distribution of galaxies or from the $\lambda$CDM
simulations of Kauffmann et~al. (1999). This may indicate some environmental
dependence on whether a galaxy emits detectable \lya\ emission.

The detection of too many voids in our data is suggestive, but further
observations with much more uniform sensitivity over much wider
fields, combined with follow-up spectroscopy, will be required to
definitively measure the topology of the distribution of high redshift
galaxies.

\acknowledgements

We thank K. Weaver and R. Mushotzky for useful discussions, and the
VIRGO consortium for making the outputs of their n-body simulations
publicly available. We thank the referee W. Keel for comments
which improved the manuscript.  This study was funded by a NASA grant
NRA--98--03--UVG--011. CTIO is operated by the Association of
Universities for Research in Astronomy (AURA), Inc., under cooperative
agreement with the National Science Foundation.

This work was supported by the STIS IDT through the National Optical
Astronomical Observatories and by the Goddard Space Flight Center.

\references

\reference{} Abell, G.~O., Corwin, H.~G., \& Olowin, R.~P.\ 1989, ApJS, 70, 1

\reference{} Adelberger, K.~L., Steidel, C.~C., Giavalisco, M., Dickinson, M.,
Pettini, M., \& Kellogg, M.\ 1998, ApJ, 505, 18

\reference{} Barger, A.~J., Cowie, L.~L.,  Mushotzky, R.~F., \&  Richards, E.~A.\  2001, AJ, 121, 662

\reference{} Bautz, M.~W., \& Garmire, G.~P.\ 2001, AAS, 199, 100.10

\reference{} Beers, T.~C., Flynn, K., \& Gebhardt, K.\ 1990, \aj, 100, 32

\reference{} Bergeron, J., Petitjean, P., Cristiani, S., Arnouts, S., Bresolin, F., \& Fasano, G.\  1999, A\&A, 343, L40

\reference{} Bertin, E., \& Arnouts, S.\ 1996, \aaps, 117, 393

\reference{} Campos, A., Yahil, A., Windhorst, R.~A., Richards, E.~A.,
Pascarelle, S., Impey, C., \& Petry, C.\ 1999, ApJL, 511, L4

\reference{} Chapman, S.~C.,  Lewis, G.~F., Scott, D., Richards, E., Borys, C., Steidel, C.~C., Adelberger, K.~L., \& Shapley, A.~E.\ 2001, ApJ, 548, L17

\reference{} Cowie, L.~L., \& Hu, E.~M.\ 1998, AJ, 115, 1319

\reference{} Ellingson, E., Yee, H.~K.~C., \& Green, R.~F.\ 1991, ApJ, 371, 49

\reference{} Fabian, A.~C., Arnaud, K.~A., Nulsen, P.~E.~J., \& Mushotzky, R.~F.\ 1986, ApJ, 305, 9

\reference{} Fall, S.~M., \& Efstathiou, G.\ 1980, MNRAS, 193, 189

\reference{} Fardal, M.~A., Katz, N., Gardner, J.~P., Hernquist, L., Weinberg, D.~H., \& Dav\'{e}, R.\  2001, ApJ, 562, 605

\reference{} Francis, P.~J., \& Hewett, P.~C.\ 1993, AJ, 105, 1633

\reference{} Francis, P.~J., et~al.\ 1996, ApJ, 457, 490

\reference{} Francis, P.~J., Woodgate, B.~E., \& Danks, A.~C.\ 1997, ApJ, 482, L25
\reference{} Francis, P.~J., Wilson, G.~M., \& Woodgate, B.~E.\ 2001, PASA, 18, 64

\reference{} Francis, P.~J., et~al.\ 2001, ApJ, 554, 1001

\reference{} Geller, M.~J., \& Huchra, J.~P.\ 1989, Science, 246, 897

\reference{} Giavalisco, M., Steidel, C.~C., Adelberger, K.~L., Dickinson, M.~E., Pettini, M., \& Kellogg, M.\ 1998, ApJ, 503, 543

\reference{} Haiman, Z.,  Spaans, M., \& Quataert, E.\ 2000, ApJ, 537, L5

\reference{} Hall, P.~B., Osmer, P.~S., Green, R.~F., Porter, A.~C., \& Warren, S.~J.\ 1996a, ApJ, 462, 614

\reference{} Hall, P.~B., Osmer, P.~S., Green, R.~F., Porter, A.~C., \& Warren, S.~J.\ 1996b, ApJ, 471, 1073

\reference{} Hawkins, M.~R.~S.\ 2000, A\&AS, 143, 465

\reference{} Hogg, D.~W.\ et~al.\ 1998, AJ, 115, 1418

\reference{} Hu, E.~M., \& McMahon, R.~G.\ 1996, Nature, 382, 281

\reference{} Jerjen, H., Bingelli, B., \& Freeman, K.~C.\ 2000, AJ, 119, 593

\reference{} Kauffmann, G., Nusser, A., \& Steinmetz, M.\ 1997, MNRAS, 286, 795

\reference{} Kauffmann, G., Colberg, J.~M., Diaferio, A., \& White, S.~D.~M.\ 1999, MNRAS ,303, 188

\reference{} Keel, W.~C., Cohen, S.~H., Windhorst, R.~A., \& Waddington, I.\ 1999, AJ, 118, 2547

\reference{} Kennicutt, R.~C., Jr.\ 1983, ApJ, 272, 54

\reference{} Krolik, J.~H., \& Begelman, M.~C.\  1986, ApJ, 308, L55

\reference{} Kron, R.~G.\ 1980, ApJS, 43, 305

\reference{} Kurk, J.~D., Pentericci, L., R{\o}ttgering, H.~J.~A., \& Miley, G.~K.\ 2001, ApSSS, 277, 543

\reference{} Landolt, A.~U.\ 1992, \aj, 104, 340

\reference{} Le F\'{e}vre, O., Deltorn, J.~N., Crampton, D., \& Dickinson, M.\ 1996, ApJ., 471, L11

\reference{} Lowenthal, J.~D., Hogan, C.~J., Green, R.~F., Caulet, A., Woodgate, B.~E., Brown, L., \& Foltz, C.~B.\ 1991, ApJ, 377, L73

\reference{} Madau, P., Pozzetti, L., \& Dickinson, M.\ 1998, ApJ, 498, 106

\reference{} Mo, H.~J., Mau, S., \& White, S.~D.~M.\ 1999, MNRAS, 304, 175

\reference{} M{\o}ller, P., \& Fynbo, J.~U.\ 2001, A\& A, 372, L57

\reference{} Mushotzky, R.~F., Cowie, L.~L., Barger, A.~J., \& Arnaud, K.~A.\ 2000, Nature, 404, 459

\reference{} Ouchi, M. et~al.\ 2003,  ApJ, 582, 60O

\reference{} Pascarelle, S.~M., Windhorst, R.~A., \& Keel, W.~C.\ 1998, AJ, 116, 2659

\reference{} Pascarelle, S.~M., Windhorst, R.~A., Keel, W.~C., \& Odewahn, S.~C.\ 1996, Nature, 383, 45

\reference{} Palunas, P., et~al.\ 2000, ApJ, 541, 61

\reference{} Pritchet, C.~J.\ 1994, PASP, 106, 1052

\reference{} Pustilnik, S.~A., Engels, D., Ugryumov, A.~V., Lipovetsky, V.~A.,
Hagen, H.-J., Kniazev, A.~Y., Izotov, Y.~I., \& Richter, G.\ 1999, A \&\ AS, 137, 299


\reference{} Rhoads, J.~E.\  2000, PASP, 112, 703

\reference{} Shapley, A.~E., Steidel, C.~C., Pettini, M., \& Adelberger, K.~L.\ 2003, ApJ, 588, 65 

\reference{} Schlegel, D.~J., Finkbeiner, D.~P., \& Davis, M.\ 1998, ApJ, 500, 525

\reference{} Shectman, S.~A., Landy, S.~D., Oemler, A., Tucker, D.~L., Huan, L., Kirshner, R.~P., \& Schechter, P.~L.\ 1996, ApJ, 470, 172

\reference{} Silk, J.~\& Weinberg, D.\ 1991, \nat, 350, 272 

\reference{} Steidel, C.~C., Giavalisco, M., Pettini, M., Dickinson, M., \& Adelberger, K.~L.\ 1996, ApJ Letters 462, L17

\reference{} Steidel, C.~C., Adelberger, K.~L., Dickinson, M., Giavalisco, M.,
Pettini, M., \& Kellogg, M.\ 1998, ApJ, 492, 428

\reference{} Steidel, C.~C., Adelberger, K.~L., Shapley, A.~E., Pettini, M., Dickinson, M., \&  Giavalisco, M.\ 2000, ApJ, 532, 170

\reference{} Taniguchi, Y., \& Shioya, Y.\ 2000, ApJ, 532, L13 

\reference{} Taniguchi, Y., Shioya, Y., \& Kakazu, Y.\ 2001, ApJ, 562, L15

\reference{} Teplitz, H.~I, Malkan, M.~A., \& McLean, I.~S.\ 1998, ApJ, 506, 519 

\reference{} Villar-Martin, M., Tadhunter, C., \& Clark, N.\  1997, A\& A, 323, 21

\reference{} White, S.~D.~M., \& Frenk, C.\ 1991, ApJ, 379, 52

\reference{} Windhorst, R.~A., Keel, W.~C., \& Pascarelle, S.~M.\ 1998, ApJ, 494, L27

\end{document}